# Nonlinear dynamics and chaos in an optomechanical beam

D. Navarro-Urrios,[1] N. E. Capuj,[2] M. F. Colombano,[1] P. D. García,[1] M. Sledzinska,[1] F. Alzina,[1] A. Griol,[3] A. Martinez,[3] C. M. Sotomayor-Torres[1,4]

[1] Catalan Institute of Nanoscience and Nanotechnology (ICN2), CSIC and The Barcelona Institute of Science and Technology, Campus UAB, Bellaterra, 08193 Barcelona, Spain
[2] Depto. Física, Universidad de la Laguna, La Laguna, Spain
[3] Nanophotonics Technology Center, Universitat Politècnica de València, Spain
[4] Catalan Institute for Research and Advances Studies ICREA, Barcelona, Spain
e-mail: daniel.navarro@icn2.cat

**Abstract:** Optical non-linearities, such as thermo-optic effects and free-carrier-dispersion, are often considered as undesired effects in silicon-based resonators and, more specifically, optomechanical (OM) cavities, affecting the relative detuning between an optical resonance and the excitation laser. However, the interplay between such mechanisms could also enable unexpected physical phenomena to be used in new applications. In the present work, we exploit those non-linearities and their intercoupling with the mechanical degrees of freedom of a silicon OM nanobeam to unveil a rich set of fundamentally different complex dynamics. By smoothly changing the parameters of the excitation laser, namely its power and wavelength, we demonstrate accurate control for activating bi-dimensional and tetra-dimensional limit-cycles, a period doubling route and chaos. In addition, by scanning the laser parameters in opposite senses we demonstrate bistability and hysteresis between bi-dimensional and tetra-dimensional limit-cycles, between different coherent mechanical states and between tetra-dimensional limit-cycles and chaos. As a result of implementing the Rosenstein algorithm to the experimental time series we have extracted a Largest Lyapunov Exponent of LLE=$1.3 \times 10^5$ s$^{-1}$, which is between one and two orders of magnitude lower than the typical oscillating frequencies of the system. Most of the experimental features can be well reproduced with a model of first-order non-linear differential equations coupled through the number of intracavity photons. Our findings open several routes towards exploiting silicon-based OM photonic crystals for systems with memory and, more specifically, for chaos based applications such as secure information transfer or sensing.

**Introduction**

The long-term solutions of whatever one- or two-dimensional non-linear system restrict to nothing more complicated than a stable/unstable fixed point or a limit cycle, the possible bifurcation types being already well studied [1]. If more than two dimensions become at play trajectories may become much more complex, eventually displaying aperiodicity and extreme sensibility to initial conditions, i.e., exhibiting chaotic dynamics. The Lorenz equations are the paradigmatic example of how complex and rich could be the solution of a relatively simple deterministic set of differential equations [2,3], where only one of the three is non-linear. Already five decades old, wide ranges of the parameters governing the equations are still unexplored. Those equations have been applied to successfully model a wide range of phenomena with a drastically different physical origin.

Classical non-linear dynamics has been explored quite extensively in different OM architectures. Indeed, several mechanisms and techniques can drive a mechanical mode into a state of large mechanical oscillation amplitude by exploiting the radiation pressure force, namely stimulated emission [4], dynamical back-action [5], forward stimulated Brillouin scattering [6] and self-pulsing [7]. Interesting features and tools can appear in that regime, e.g. the existence of several stable mechanical-amplitude solutions for a fixed set of external parameters, which was demonstrated experimentally in refs. [8,9]. Moreover, for sufficiently high laser power, the system could eventually enter into a chaotic regime, which may be exploited for chaos-based secure data communication or sensing, among other applications. In this regard, integrated OM systems may present advantages with respect to coupled lasers [10] or hybrid optoelectric oscillators [11] in terms of the ease of integration, scalability and engineering of the nonlinearities [12]. Despite their potential, only the works of Carmon et al. [13, 14] and Monifi et al. in dielectric microtoroids [15] and by Wu et al. [16] and our group [17] in silicon (Si) based integrated devices have addressed chaos in integrated OM architectures.

In this manuscript, we tackle the previous issues in a silicon-based one-dimensional OM photonic crystal in which the physical magnitudes governing its non-linear dynamics are of very diverse origin, namely temperature, free-carrier population and mechanical deformation. Those intercouple through the number of intracavity photons, which affects and is affected by the previous magnitudes. In contrast to the well-known static fixed points of an OM system [18], we present here a heterogeneous variety of stable dynamical solutions that, in some specific cases, coexist, giving rise to bi-stability and hysteresis. In particular, we report and carefully characterize radio-frequency (RF) spectra and temporal series of the optical transmission in a six-dimensional chaotic regime.

**General properties of the sample**

The device presented here is a one dimensional OM photonic crystal fabricated using standard Si nanofabrication tools (see Supplementary Discussion 1) on a Silicon-On-Insulator (SOI) wafer. As seen from the SEM top-view of Fig. 1a, the crystal lattice constant is quadratically reduced towards the center of the beam, hereby defining high-Q optical cavity modes.

In order to accurately model the fabricated OM system and account for the differences with respect to the nominal geometry, the in-plane geometry is imported from the SEM micrograph (Fig. 1a) into the FEM solver, the thickness being that of the SOI wafer. This procedure provides a good agreement between the measured optical modes and those extracted from simulations (Fig. 1b).

In the following, we investigate the third optical mode (Fig. 1c) of the OM photonic crystal. It displays an asymmetric field distribution with respect to the xz plane, giving rise to surprisingly high values of the single-particle OM coupling rate ($g_{o,OM}$, red dots of Fig. 4d) for in-plane (xy plane) flexural modes with an antinode in the beam center. In Fig. 4d, we have highlighted the one having three antinodes along the x-direction ($\Omega_m$=54 MHz) because of being at the heart of the complex dynamics revealed by the system. The RF signal associated to optical transduction of thermally driven mechanical eigenmodes is also reported in Fig, 4d, whose peaks positions show a good agreement with the modes predicted to display high $g_{o,OM}$ values.

An evident discrepancy between the FEM simulations of $g_{o,OM}$ and the actual transduced RF signal concerns out-of-plane flexural modes, which should not present a significant $g_{o,OM}$ due to symmetry considerations. One of the most likely reasons for the opposite is that geometric inhomogeneities along the z axis may break the symmetry along that axis, unbalancing the contributions from the top and bottom Si-air interfaces. As it will be discussed later on, the fundamental out-of-plane flexural mode ($\Omega_m'$=5 MHz, first peak in the RF spectrum) plays a key role on the chaotic dynamics of the system.

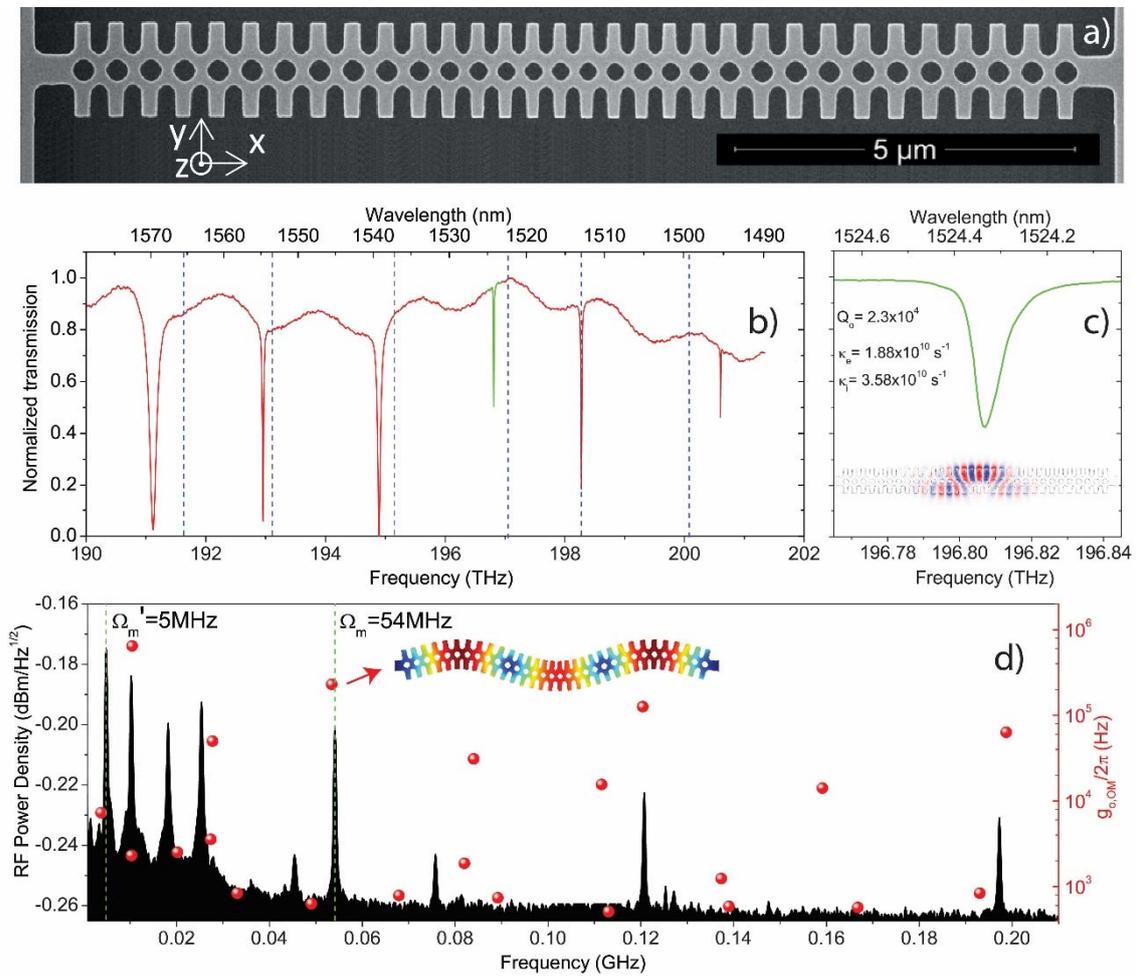

*Figure 1.* *a) SEM micrograph of the fabricated Si OM photonic crystal. b) Normalized transmission spectrum recorded for low laser power (<10μW) c) Zoom of the optical resonance under study. d) Left axis. RF spectrum obtained by exciting the cavity with the first-order optical mode. Right axis. Calculated values of $g_{o,OM}$ for each of the mechanical eigenmodes are reported with red dots. The deformation profile of the nanobeam associated to the $\Omega_m$=54 MHz mode is also shown.*

**Pure self-pulsing and coupled self-pulsing/phonon lasing: bistability and hysteresis**

It is well known that large energies stored in optical resonators may give rise to strong non-linear effects. The two most intense sources of optical non-linearities in Si-based optical resonators are: thermo-optic (TO) effects, which red-shifts the resonance position ($\lambda_r$) proportionally to the average cavity temperature increase (Δ$T$); and free carrier dispersion (FCD), which induces a blue-shift proportional to the the free carrier population (N). The dynamics of Δ$T$ and $N$ can be described by a system of rate equations (see Supplementary Discussion 4) that are coupled through the number of photons within the cavity:

$$n = n_o \frac{\Delta\lambda_o^2}{4(\lambda_l - \lambda_r)^2 + \Delta\lambda_o^2}; \qquad (1)$$

where $n = n_o = 2 P_l \kappa_e \lambda_o / \kappa^2 hc$ in perfect resonance. $P_l$ and $\lambda_l$ are the laser power and wavelength, respectively; $\kappa_e$ and $\kappa$ and are the extrinsic and overall optical damping rates, respectively, the latter determining the cavity resonant linewidth ($\Delta\lambda_o = \lambda_o^2 \kappa / 2\pi$). The only two parameters that have been modified experimentally within this work are $P_l$ and $\lambda_l$ with a resolution below 1 µW and 1 pm, respectively. It is also worth noting the possibility of tailoring, in a much less accurate way, other parameters entering the equations, such as $\kappa$, $\kappa_e$ and the heat dissipation rate, by adjusting the fiber/sample relative alignment.

In most of the situations, the long term dynamic solution of the {$\Delta T$, $N$} system is a stable fixed point. However, for specific combinations of $P_l$ and $\lambda_l$, the fixed point undergoes a supercritical Hopf bifurcation, transmuting into an unstable fixed point surrounded by a stable limit cycle. The long-term solution of the {$\Delta T$, $N$} system is then that limit-cycle, termed from now on self-pulsing (SP) [19], where the cavity resonance oscillates periodically around the laser at a frequency denoted by $\nu_{SP}$. The total time required to complete the SP limit-cycle is thus $1/\nu_{SP}$, although it is not necessarily drawn at a constant pace. When the SP limit-cycle is active, light within the cavity gets modulated in a strongly anharmonic way, creating an "optical frequency comb" with multiple peaks spectrally located at integers of $\nu_{SP}$ whose intensity decreases with frequency. An important consequence is that the intracavity radiation pressure force ($F_o$), which is proportional to $n$, gets modulated in an equivalent fashion. In a previous work [7], we demonstrated that if the $M^{th}$ harmonic of $F_o$ has a significant overlap with the linewidth of a mechanical mode with sufficient $g_{o,OM}$, the latter enters a high-amplitude, coherent regime, which fulfills the requirements to be identified as "phonon lasing" [20,21].

The transition from pure SP dynamics to the coupled SP/phonon lasing counterpart is effectively that from a bi-dimensional to a tetra-dimensional coupled non-linear system. Indeed, the mechanical oscillator is not just being unidirectionally driven, since the OM coupling provides an effective back coupling that forces $\nu_{SP}$ to be frequency-locked to $\nu_{SP}=\Omega_m/M$ over a wide spectral range of the laser source. Lower M values result in stronger $F_o$ and higher coupling strengths between the SP and the mechanical coherent oscillator. Once the system is found in a tetra-dimensional limit cycle it becomes very robust, a characteristic that can be quantified at some extent from the spectral width of the frequency-locked plateau and the low phase noise of the RF signal. These features are a direct indication of ranges of laser parameters in which there coexist two stable states of the system, namely the bi-dimensional SP and the tetra-dimensional SP/phonon lasing. The existence of bistability enables the possibility of frequency-jumps and hysteresis as the laser parameters are scanned. To check for these features we have studied the full RF spectra as a function of $\lambda_l$ at $P_l$=2 mW. The experimental data have been taken by sweeping $\lambda_l$ from bottom to top (Fig. 2a) and viceversa (Fig. 2b). As expected, $\lambda_l$-frequency-jumps and $\lambda_l$-hysteretic regions appear where the system undergoes state transitions. By singling out the position of the first RF harmonic (Figs. 2c) the $\lambda_l$-bistable regions are clearly exposed.

Most of the $\lambda_l$-bistable features described above are well reproduced by numerically solving the four-dimensional coupled non-linear system {$\Delta T$, $N$, $u$, $du/dt$} (see Supplementary Discussion 5), where *u* and *du/dt* are the generalized coordinate for the displacement of the mechanical mode and its derivative, respectively. An abrupt transition between two available states occurs if the frequency difference between them ($\Delta\nu$) reaches a specific amount $\Delta\nu<\Delta\nu_{up}$ ($\Delta\nu>\Delta\nu_{down}$) for an up (down) transition, where $\Delta\nu_{up}<\Delta\nu_{down}$. In order to unveil a hysteresis cycle, $\Delta\nu$ must fulfil the conditions for performing each of the transitions along the explored parameter path. In general, in our system we observe that $\Delta\nu_{up}$ and $\Delta\nu_{down}$ increase (i.e., the frequency-jumps are larger) with $P_l$ and for lower M values.

$\lambda_l$-bistable regions unfold experimentally and numerically solving the model at different spectral windows. It is worth noting that it is not possible to simultaneously operate within two of these windows because of being associated to the same photonic resonance. However, it is viable to probe, taking care of not inducing further optical non-linearities, photonic resonances alternative to the one being pumped. In that case, the dynamics of the probe transmitted signals would be coherent with that of the pump, since each one just follows the common dynamics of the effective refractive indices [15].

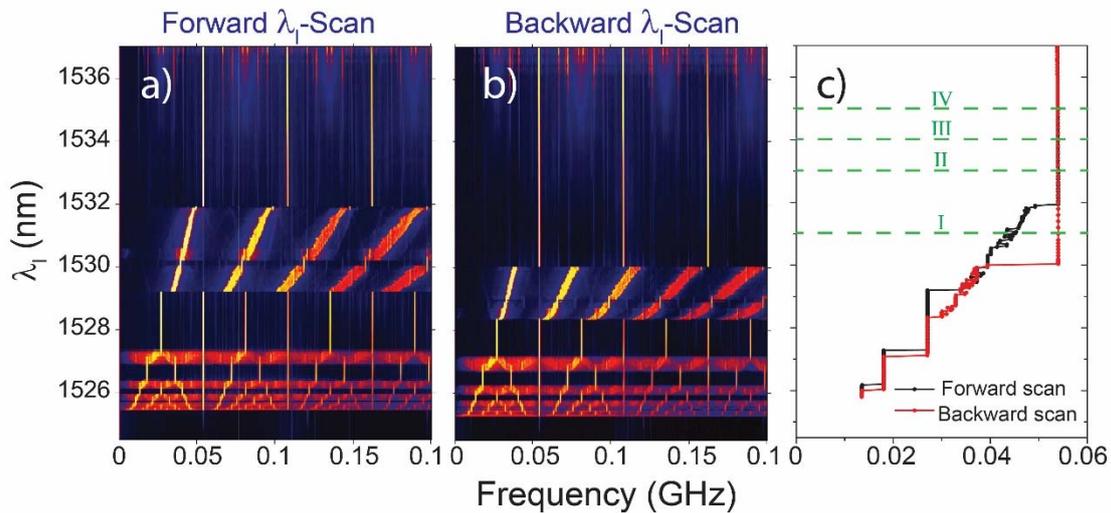

*Figure 2. Panels **a)**-**b)**. Color contour plot of the RF spectrum as a function of $\lambda_l$ obtained at $P_l$=2 mW. Panel **a)** correspond to a forward $\lambda_l$-scan while panel **b)** to a backward $\lambda_l$-scan. Panel **c)** plots the spectral position of the first RF harmonic. Horizontal dashed lines in panel c) indicate positions in which a $P_l$ scan has been performed (see Figure 3).*

The laser parameters space has been further explored by monitoring the system response to changes on $P_l$ for fixed values of $\lambda_l$. The analysed domain lies in the vicinity of the widest bistable region of Fig. 2c (the precise $\lambda_l$ under analysis are highlighted with dashed lines). The results are reported in Fig. 3 by plotting the spectral position of the first RF harmonic.

$P_l$-bistability and $P_l$-hysteresis appear only if $\lambda_l$ is above the $\lambda_l$-frequency-jump of the forward $\lambda_l$ scan. If $\lambda_l$ lies well within the $\lambda_l$-bistable region, such as in I, the system can start from either of the two stable states showed in Fig. 3 (red and black curve). However, within the studied $P_l$ range along I, $\Delta\nu$ only satisfies the requirements for performing a down-transition ($\Delta\nu>\Delta\nu_{down}$), being unable to accomplish an up-transition to a M=1 state. In order to drive the system to a M=1 state, $\lambda_l$ has to be increased and, subsequently, decreased.

A $P_l$-bistable region unwraps for $\lambda_l$ values in the close vicinity of the $\lambda_l$-frequency-jump, whose width decreases with $\lambda_l$ (see green and blues curves at II and III). It fully closes at IV, where the system stays robustly in the M=1 plateau. Interestingly, $P_l$-bistability involves two different tetra-dimensional SP/phonon lasing states, namely the pervasive $\Omega_m$=54 MHz mode with M=1, and a 198 MHz mode (associated to an in-plane flexural mode having 7 antinodes along the x-direction, see Fig. 1d and Supplementary Discussion 2) with M=4 (i.e., the main RF peak appears at 49.5 MHz).

State-switching within the $P_l$-bistable region would be enabled straightforwardly by applying power pulses/dark pulses), the pulse height needed being dependent on the specific $\lambda_l$ (ranging between few μW and several hundreds of μW). The mechanical lifetime and frequency of the modes involved would currently limit switching speeds to the MHz range, but this could be further improved by pushing the SP frequencies to the GHz range with a smart tailoring of the heat dissipation and free-carrier recombination rates.

Regions of the laser parameters space supporting more than two stable states could also be present. Indeed, it is plausible that the $P_l$-bistable region reported in Fig. 3 could support three stable states, namely two different tetra-dimensional SP/phonon lasing states and bi-dimensional SP. Unfortunately, we have been unable to reach all of them by just scanning $P_l$.

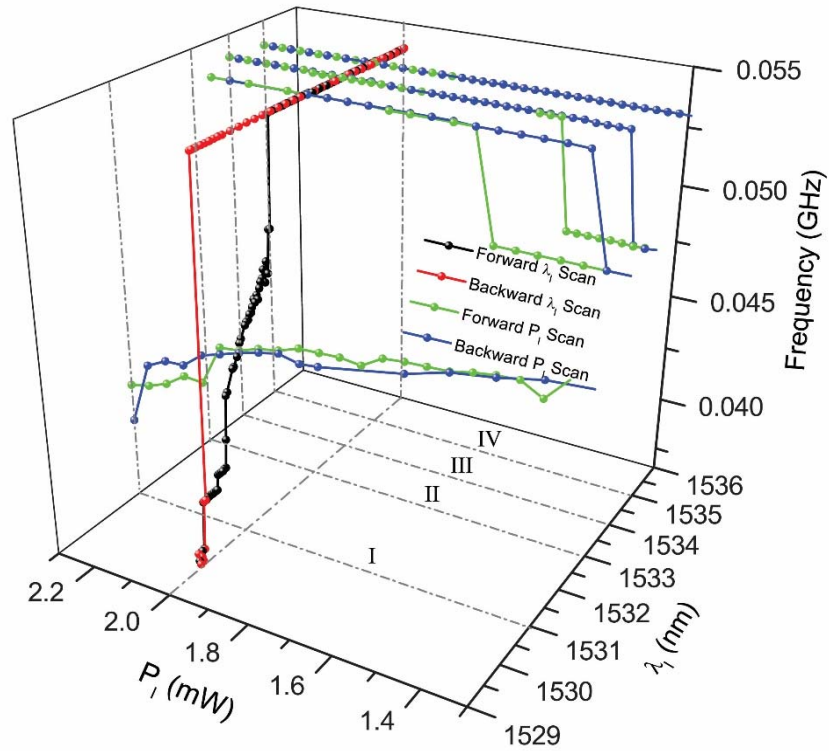

**Figure 3.** *Spectral position of the first RF harmonic as a function of $P_l$ and $\lambda_l$.*

**Non-linear dynamics at high $P_l$. Period doubling bifurcation of cycles and onset of chaos: bistability and hysteresis**

The dynamical solution of the system further increases its complexity for specific ranges of $P_l$ and $\lambda_l$, giving rise to period-doubling cycles and the onset of chaos. For continuous dynamical systems the Poincaré–Bendixson theorem states that period doubling bifurcation of cycles or bifurcations towards strange attractors can only arise in three or more dimensions [1]. Therefore, this kind of bifurcations can occur, in principle, if the OM device is found in a tetra-dimensional SP/phonon lasing state and $P_l$ and $\lambda_l$ are varied from there. Indeed, for high enough $P_l$ values, new complex dynamics appear at the sides of the plateaus. Fig. 4 illustrates the most striking case, covering both sides of the M=1 plateau while pumping at $P_l$=4 mW. Here, the transition between the M=2 and M=1 plateaus does not involve pure SP states, contrary to what observed at $P_l$=2 mW (see Fig. 2). Indeed, along the transition, the dimension of the system is, at least, four, following a route in which the attractor undergoes subsequent period-doubling bifurcations (see lower part of Figs. 4a and 4b and Fig. 4c). A much broader linewidth of the $\Omega_m/2$ peak of the M=1 first period doubling enables its discrimination from the M=2 coherent state (red and blue curves of the inset of Fig. 4c). Subsequent period doubling bifurcations appear before entering into the M=1 coherent regime, whose peak intensity decreases as the period doubling level increases.

On the other side of the M=1 plateau the system suffers another bifurcation in which the system abruptly enters into a chaotic regime within a strange attractor volume in the phase space. The RF spectra below and above that bifurcation are shown in Fig. 4d (black and red curves, respectively). A chaotic behavior is also foreseen by our numerical simulations for high enough $\lambda_l$ values at high $P_l$ (see Supplementary Discussion 6 and 7). The simplest version of our model, which considers a single mechanical mode at $\Omega_m$, predicts: the onset of chaos in $\Delta T$ and $N$ by a smooth period-doubling kind of route and; a coherent high-amplitude oscillation of the mechanical mode. This would lead to a flat broad-band RF spectrum with sharp peaks only at harmonics of $\Omega_m$. In the experiment, the onset of chaos is abrupt and leads to a broad-band RF spectrum as well, but with a much richer structure of sharp peaks (red curve of Fig. 4d). The most intense ones are associated to the $\Omega_m$ mode (main peak plus harmonics) that was lasing in the plateau, similarly to what reported in the numerical simulation (see Supplementary Discussions 6 and 7). The remaining peaks are associated to the activation of the fundamental out-of-plane flexural mode described before ($\Omega_m'$=5 MHz), providing peaks at $\Omega_m'$, at its harmonics and at sidebands of the $\Omega_m$ associated peaks. On the basis of these experimental results we interpret that, as predicted, a chaotic regime is established out of the SP/single mechanical mode lasing state, thus creating a broad band $F_o$. However, this is not the stable solution of the system since, in response to $F_o$, other mechanical modes displaying large enough $g_{o,OM}$ can be activated. The non-linear system thus increases its dimension by an amount equal to twice the number of mechanical modes involved. We have confirmed this hypothesis by including a second harmonic oscillator at $\Omega_m'$ in our model (see Supplementary Discussion 7) driven by $F_o$, the result being that the chaotic dynamics is only present in $\Delta T$ and $N$, while the two mechanical modes oscillate coherently at incommensurate frequencies. It is noteworthy that the only additional mode to be amplified is an out-of-plane flexural one having a large $g_{o,OM}$ value (see Fig. 1d), the other in-plane ones, although having large $g_{o,OM}$ values as well, are probably damped because of mechanical mode competition [22].

There are other two remarkable features reported in Figs. 4a and 4c: i) the absence of hysteresis in the transition between M=2 and M=1 and ii) the hysteresis cycle involving chaos and the SP/single mode lasing limit-cycle. The former contrasts with what reported in Fig. 2, leading us to conclude that hysteresis only appears when the transition is through bi-dimensional SP, i.e., at low $P_l$. The observation of ii) is an indication of two types of coexisting attractors: fixed points and strange attractors. It also means that a large enough instantaneous perturbation can knock the coherent oscillation into chaos or viceversa. The complete insight of the latter bistability exceed the scope of this work, but we believe it is a direct consequence of the coupling between the two mechanical modes through *n*.

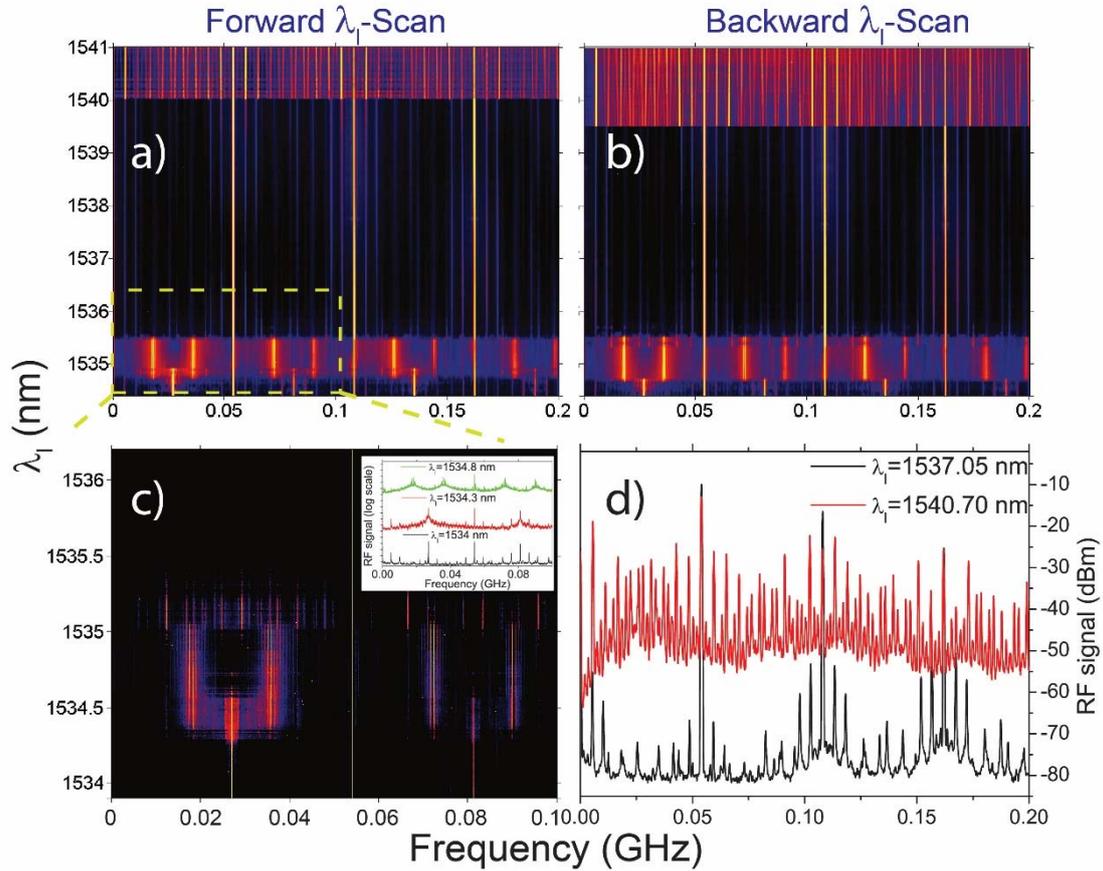

*Figure 4.* Panels **a)** and **b)**. Color contour plot of the RF spectrum as a function of $\lambda_l$ obtained at $P_l$=4mW. Panels **a)** correspond to a $\lambda_l$-forward scan while panel **b)** to $\lambda_l$-backward scan. Panel **c)** displays a zoom of the transition between M=2 to M=1 plateaus through subsequent period doubling bifurcations. The color contrast has been altered with respect to panel c) to highlight the period doubling route. The **inset** of panel c) shows RF spectra at three different wavelengths within the M=2 to M=1 transition. Panel **d)** shows RF spectra below and above the bifurcation onsetting chaos (black and red curves respectively).

The time evolution of the transmitted signal below and above the abrupt bifurcation delimiting the onset of chaos is reported in Fig. 5. As expected, above the bifurcation, the signal is aperiodic while below the bifurcation the signal is coherent. Figs. 5b and 5d plot the transmitted signal vs itself delayed 10 ns and itself delayed 20 ns, aiming to reconstruct the embedding state in a projection on a three-dimensional space. A limit cycle is reconstructed for the coherent case, while the described trajectory in the chaotic case tends to fill a restricted volume.

We have finally applied the Rosenstein algorithm [23] to 10 μs-long transmission registers taken with a temporal resolution of 10 ps (see Supplementary Discussion 8). The output is the temporal evolution of the average divergence of initially close state-space trajectories (Fig. 5e). The applied algorithm starts converging for an embedding dimension (m) m≥6. The latter must be at least equal to the dimension of the system [23], which agrees with the dimension of our model considering two mechanical modes. Largest Lyapunov Exponents (LLE) quantify the exponential behavior of the divergence and estimate the amount of chaos in the system. The sensitive dependence on initial conditions, quantified by LLE, preclude long-term predictions, but it promises improved short-term predictions [24]. From the exponential fit of the blue curve of Fig. 5e we extract an LLE value of $1.3 \times 10^5$ s$^{-1}$. This value is about one order of magnitude lower than $\Omega_m'$ and two orders of magnitude lower than $\Omega_m$, which are the typical frequencies of the system. Obviously, the divergence growing trend saturates at some point, since trajectories cannot get any further than the scale of the attractor volume, which is on the order of the amplitude of the transmission oscillations. Aiming to check for the validity of the used parameters we have applied the same algorithm to the coherent signal (red curve of Fig. 5e), the result being that the trajectories stay parallel at a divergence value almost an order of magnitude shorter than in the chaotic case.

Notice that it was not possible for us to compare the obtained LLE values to those reported elsewhere for OM integrated systems: in Ref. [15], those are reported without units; in Ref. [16], LLE values are much greater than the typical frequencies of the system, i.e., the horizon of predictability is much shorter than the time required to perform a single oscillation.

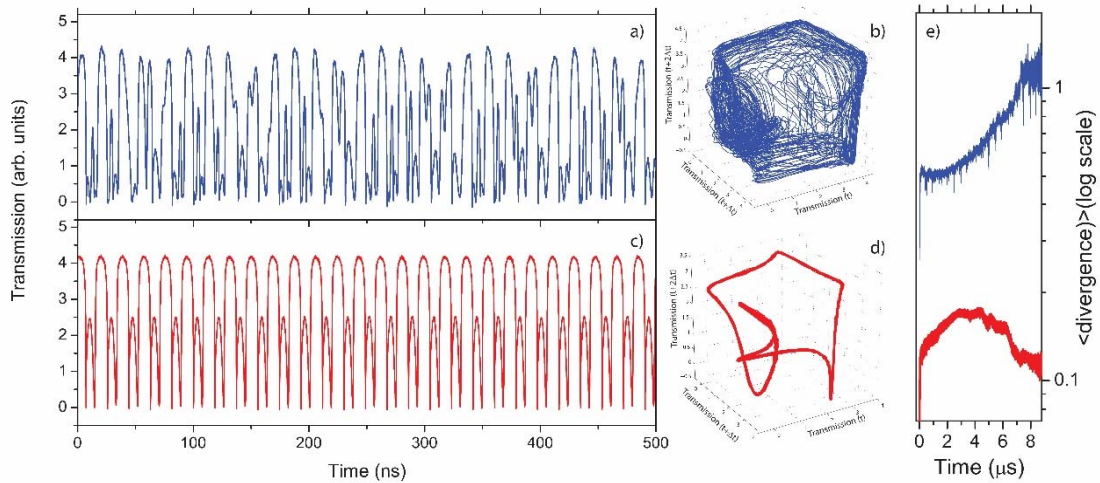

*Figure 5. Time evolution of the transmitted signal above and below the bifurcation onsetting chaos (panels a) and c) respectively). Reconstruction of the embedding state in a three-dimensional projection as extracted from the temporal signals (panels b) and d)). Panel e) reports the time evolution of the divergence (in log scale) extracted from the experimental signals by applying Rosenstein algorithm. The input parameters are: m=8; delay=10ns; mean period=20ns.*

**Conclusions and outlook**

We have reported a prolific set of complex dynamical solutions, including chaos, observed in a single and compact Si-based OM nanobeam, optically pumped with a continuous-wave, low-power laser source. The reported features are very robust and stable for hours, in spite of being achieved at atmospheric conditions of temperature and pressure.

We have provided a glimpse of the dynamical map of the system by exploring the laser parameters (power and wavelength) in different senses. In particular, at low laser powers, we have demonstrated hysteresis and bistability of two kinds involving: a bi-dimensional SP and a tetra-dimensional SP/phonon lasing state and; two tetra-dimensional SP/phonon lasing states involving two mechanical modes. The potential application, e.g. for OM memories, of the reported $P_l$-bistability is probably higher than the $\lambda_l$-bistability, since state-switching would be achieved by applying power pulses.

At high laser powers, we have observed that the transition between two consecutive tetra-dimensional SP/phonon lasing states is through a period-doubling route. In addition, we have demonstrated the onset of a six-dimensional chaos displaying bi-stability and hysteresis with a tetra-dimensional SP/phonon lasing state. We envision the use of many of those chaotic optical sources on a SOI chip in a multichannel approach. Each device will eventually show a chaotic dynamics that is dependent on the specific geometry of the OM crystal and of the parameters of the laser used to pump the optical resonance, effectively behaving as a multichannel chaotic source that could be multiplexed into a single optical fiber.

Besides the specific applications that some of the previous features may find, namely for memories, switches, sensing or secure data transmissions, the straightforward access to a large part of the parameter space makes natural the exploitation of this device as a test bed for investigating uncommon complex non-linear dynamics such as static and dynamic bifurcations, different routes to chaos, etc. The non-linear system can easily increase its complexity, for instance, by pumping several optical modes simultaneously or by weakly coupling several self-sustained OM devices, from pairs to arrays.

**Acknowledgements**: This work was supported by the European Comission project PHENOMEN (H2020-EU-713450), the Spanish Severo Ochoa Excellence program and the MINECO project PHENTOM (FIS2015-70862-P). DNU, PDG and MFC gratefully acknowledge the support of a Ramón y Cajal postdoctoral fellowship (RYC-2014-15392), a Beatriu de Pinos postdoctoral fellowship (BP-DGR 2015 (B)) and a Severo Ochoa studentship, respectively. We would like to acknowledge Jose C. Sabina de Lis, J. M. Plata Suárez, A. Trifonova and C. Masoller for fruitful discussions.

**Methods:**

**OM crystal geometry**. The unit-cell contains a hole in the central part and two symmetric stubs on the sides. Localized photonic modes appear owing to a tapered region created along 12 cells by decreasing quadratically the pitch, the radius of the hole and the stub length towards the center, with a reduction up to 83% of the values at the extremes. Outside of the tapered region, the cell is repeated 10 times on each side, so that the full length of the structure is about 15 μm. The whole beam is free-standing and anchored to the rest of the Si layer from the two extremes, forming a stripe clamped at both ends suitable for optical actuation.

**Measurement**. A tapered fiber is placed nearly parallel to the OM crystal in order to excite its localized optical modes. A tunable laser source is coupled to the fiber input and the transmitted signal is detected at the output in two ways. A Si photodiode is used for the optical spectral measurements and an InGaAs fast photoreceiver (PR) with a bandwidth of 12 GHz for the temporal and RF measurements. The RF spectra are taken using a signal analyzer with a bandwidth of 13.5 GHz. The temporal signals are taken with a 4GHz bandwidth oscilloscope. The whole setup operates at atmospheric conditions of temperature and pressure (see Supplementary Discussion 3 for more details).

# Supplementary information:

# Nonlinear dynamics and chaos in an optomechanical beam

D. Navarro-Urrios,[1] N. E. Capuj,[2] M. F. Colombano,[1] P. D. García,[1] M. Sledzinska,[1] F. Alzina,[1] A. Griol,[3] A. Martinez,[3] C. M. Sotomayor-Torres[1,4]

[1] *Catalan Institute of Nanoscience and Nanotechnology (ICN2), CSIC and The Barcelona Institute of Science and Technology, Campus UAB, Bellaterra, 08193 Barcelona, Spain*
[2] *Depto. Física, Universidad de la Laguna, La Laguna, Spain*
[3] *Nanophotonics Technology Center, Universitat Politècnica de València, Spain*
[4] *Catalan Institute for Research and Advances Studies ICREA, Barcelona, Spain*
e-mail: daniel.navarro@icn2.cat

## Materials and methods

### S1. Devices

The investigated device is an optomechanical (OM) photonic crystal whose unit-cell contains a hole in the middle and two symmetric stubs on the sides (Fig. S1). The peculiarity of this geometry resides in having a full phononic band-gap at ~4GHz (*1*). The investigation of high frequency mechanical modes is reported elsewhere *(2)*.

We describe here the geometrical parameters of our device. In a defect region consisting of 12 central cells the pitch (a), the radius of the hole (r) and the stubs length (d) are decreased in a quadratic way towards the center. The maximum reduction of the parameters is denoted by $\Gamma$. At both sides of the defect region a 10 period mirror is included. The nominal geometrical values of the cells of the mirror are a=500nm, r=150nm, and d=250nm. The total number of cells is 32 and the whole device length is about 15μm. Fig. S1 shows a SEM micrograph of one of the fabricated OM photonic crystals.

We have fabricated a set of devices in which $\Gamma$ has been varied from $\Gamma$=64% to $\Gamma$=83% of the original values. All the results presented in this work correspond to the structure with $\Gamma$=83%, but the same effects have been observed for other values of $\Gamma$.

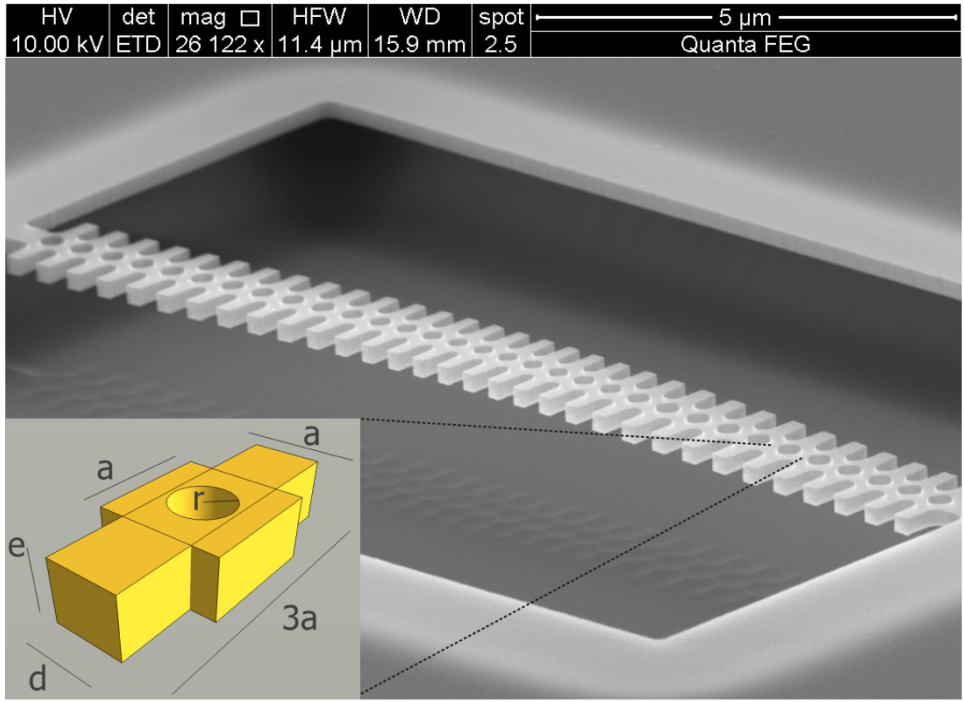

*Figure S1.* SEM micrograph of a $\Gamma$=83% OM photonic crystal. A sketch of the unit-cell is shown in the inset.

The devices were fabricated in standard silicon-on-insulator (SOI) SOITEC wafers with silicon layer thickness of 220 nm (resistivity $\rho$ ~1-10 $\Omega$ cm$^{-1}$, p-doping of ~$10^{15}$ cm$^{-3}$) and a buried oxide layer thickness of 2 µm. The pattern was written by electron beam in a 100 nm thick poly-methyl-methacrylate (PMMA) resist film and transferred into silicon by Reactive Ion Etching (RIE). Application of BHF removed the buried oxide layer and released the beam structures.

### S2. FEM simulations and OM coupling calculations

COMSOL finite-element-method (FEM) simulations of a complete structure are used to determine the fundamental cavity mode frequencies, effective masses of the mechanical eigenmodes and single-particle OM coupling rates $g_{o,OM}$.

To model the OM crystal system, the geometry of the as-fabricated structure has been measured with Scanning Electron Microscopy (SEM). The contour of the SEM image has been extracted graphically and then imported in the FEM solver (Fig. S2).

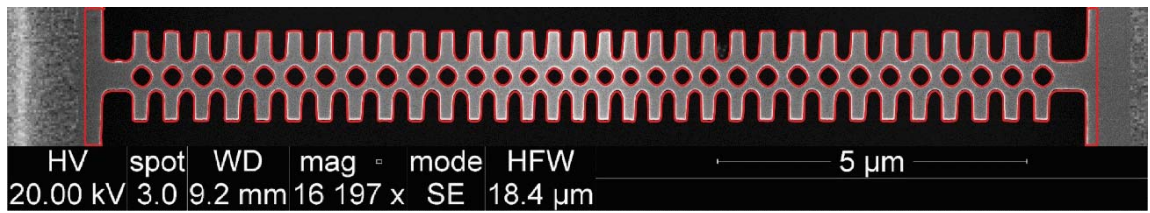

*Figure S2.* Top-view SEM micrograph of the OM system. The extracted geometrical contour imported by the FEM solver is depicted in red.

Among the many optical and mechanical modes supported by the OM crystal, there are specifically one optical and three mechanical that are discussed along the main text. Those are the third optical mode appearing at 197 THz (Fig. S3a) and the mechanical modes appearing at $\Omega_m'$=5 MHz, $\Omega_m$=54 MHz and 198 MHz (Figs. S2b, S2c and S2d, respectively), whose effective masses are 12x10$^{-12}$ g, 13x10$^{-12}$ and 11x10$^{-12}$ pg respectively.

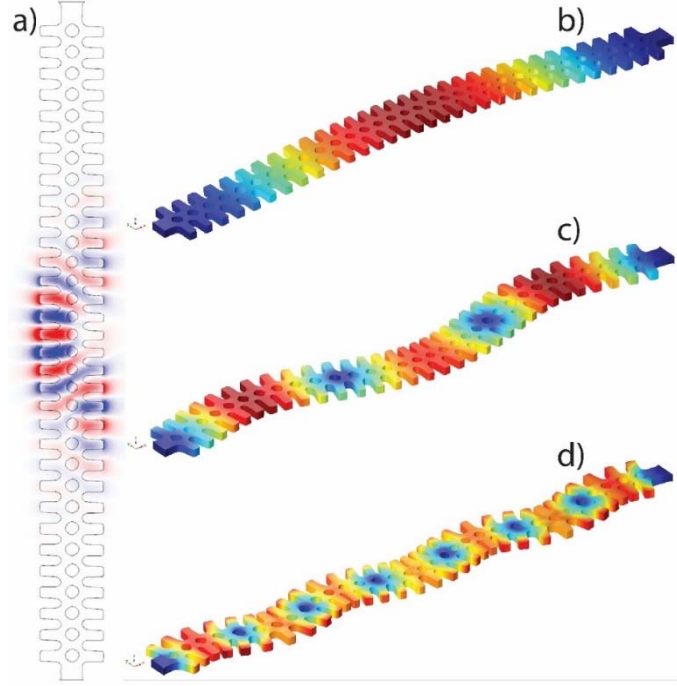

**Figure S3.** Panel a) Normalized optical $E_y$ field of the third optical mode supported by the OM crystal. Panel b), c) and d). Normalized mechanical displacement field |Q| of the $\Omega_m'$=5 MHz, $\Omega_m$=54 MHz and 198 MHz mechanical modes.

Single-particle OM coupling rates ($g_{o,OM}$) between optical and mechanical modes are estimated by taking into account both the photoelastic (PE) and the moving interfaces (MI) effects (3),(4),(5). The PE effect is a result of the acoustic strain within bulk silicon while the MI mechanism comes from the dielectric permittivity variation at the boundaries associated with the deformation.

The calculation of the MI coupling coefficient $g_{MI}$ is performed using the integral given by Johnson et al. (3);

$$g_{MI} = -\frac{\pi \lambda_r}{c} \frac{\oint (\mathbf{Q}\cdot\hat{\mathbf{n}})(\Delta\varepsilon \mathbf{E}_\parallel^2 - \Delta\varepsilon^{-1}\mathbf{D}_\perp^2)dS}{\int \mathbf{E}\cdot\mathbf{D}dV}\sqrt{\hbar/2m_{eff}\Omega_m} \qquad (S1)$$

where **Q** is the normalized displacement (max{|**Q**|}=1), $\hat{\mathbf{n}}$ is the normal at the boundary (pointing outward), **E** is the electric field and **D** the electric displacement field. $\varepsilon$ is the dielectric permittivity, $\Delta\varepsilon = \varepsilon_{silicon} - \varepsilon_{air}$, $\Delta\varepsilon^{-1} = \varepsilon^{-1}_{silicon} - \varepsilon^{-1}_{air}$. $\lambda_r$ is the optical resonance wavelength, $c$ is the speed of light in vacuum, $\hbar$ is the reduced Planck constant, $m_{eff}$ is the effective mass of the mechanical mode and $\Omega_m$ is the mechanical mode eigenfrequency, so that $\sqrt{\hbar/2m_{eff}\Omega_m}$ is the zero-point motion of the resonator.

A similar result can be derived for the PE contribution (4),(5):

$$g_{PE} = -\frac{\pi \lambda_r}{c} \frac{\langle E|\delta\varepsilon|E\rangle}{\int \mathbf{E}\cdot\mathbf{D}\,dV} \sqrt{\hbar/2m_{eff}\Omega_m} \qquad (S2)$$

where $\delta\varepsilon_{ij}=\varepsilon_{air}\,n^4 p_{ijkl} S_{kl}$, being $p_{ijkl}$ the PE tensor components, $n$ the refractive index of silicon, and $S_{kl}$ the strain tensor components.

The addition of both contributions results in the overall single-particle OM coupling rate:

$$g_{o,OM} = g_{MI} + g_{PE} \qquad (S3)$$

It is worth noting that the string-like modes studied in this work have very low PE contribution in comparison to the MI counterpart.

In Fig S4 we illustrate the MI surface density (the integrand of Eq. S1) associated to the $\Omega_m'$=5 MHz mode. As it is discussed briefly in the main text, the contribution provided by the top and bottom Si-air interfaces are both quite large, very similar in absolute value (about $g_{MI}/2\pi$=1MHz) but showing opposite signs. As a consequence of the later, a rather low value of $g_{MI}$, namely $g_{MI}/2\pi$= 7kHz, is calculated. However, geometric inhomogeneities in the real structure along the z axis may break the symmetry along that axis, unbalancing the contributions from the top and bottom Si-air interfaces.

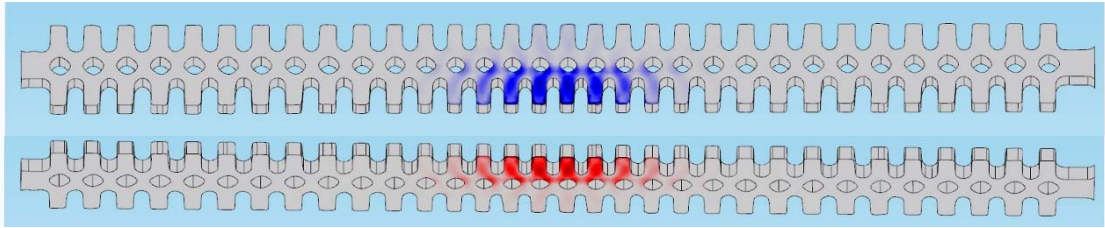

***Figure S4.*** *Normalized surface density of the integrand in Eq. S1 for the $\Omega_m'$=5 MHz mechanical mode, showing the contributions to $g_{MI}$ of the top and bottom Air-Si interfaces (Top and bottom panels respectively).*

Fig. S5 illustrates the MI surface density to the $\Omega_m$=54 MHz mode. The main contribution comes from the lateral Si-air interface. The outer stub surface and the inner hole surface contribute with different signs, but the overall value is rather large, namely $g_{MI}/2\pi$= 650 kHz.

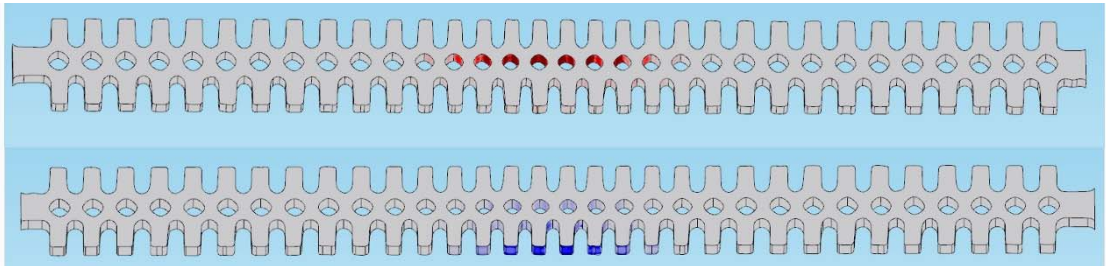

***Figure S5.*** *Normalized surface density of the integrand in Eq. S1 for the $\Omega_m$=54 MHz mechanical mode, showing the contributions to $g_{MI}$ while observing the structure from the top and from the bottom (Top and bottom panels respectively).*

## S3. Experimental set-up

The experiments are performed in a standard set-up for characterizing optical and mechanical properties of OM devices.

A tunable infrared laser covering the spectral range between 1460-1580nm is connected to a tapered fiber. The polarization state of the light entering the tapered region is set with a polarization controller. The thinnest part of the tapered fiber is placed parallel to the OM photonic crystal, in contact with an edge of the etched frame (top right photo of Fig. S6a, Fig. S6b). The gap between the fiber and the structure is about 200 nm, as roughly extracted from geometrical considerations using the radius of the fiber loop and the contact point position. A polarization analyzer is placed after the tapered fiber region.

The long tail of the evanescent field and the relatively high spatial resolution (~5 µm$^2$) of the tapered fiber locally excited the resonant optical modes of the OM photonic crystal. Once in resonance, the mechanical motion activated by the thermal Langevin force causes the transmitted intensity to be modulated around the static value (Fig. S6c).

To check for the presence of a radiofrequency (RF) modulation of the transmitted an InGaAs fast photoreceiver with a bandwidth of 12 GHz was used. The RF voltage is connected to the 50 Ohm input impedance of a signal analyzer with a bandwidth of 13.5 GHz.

All the measurements were performed in an anti-vibration cage at atmospheric conditions of air pressure and temperature.

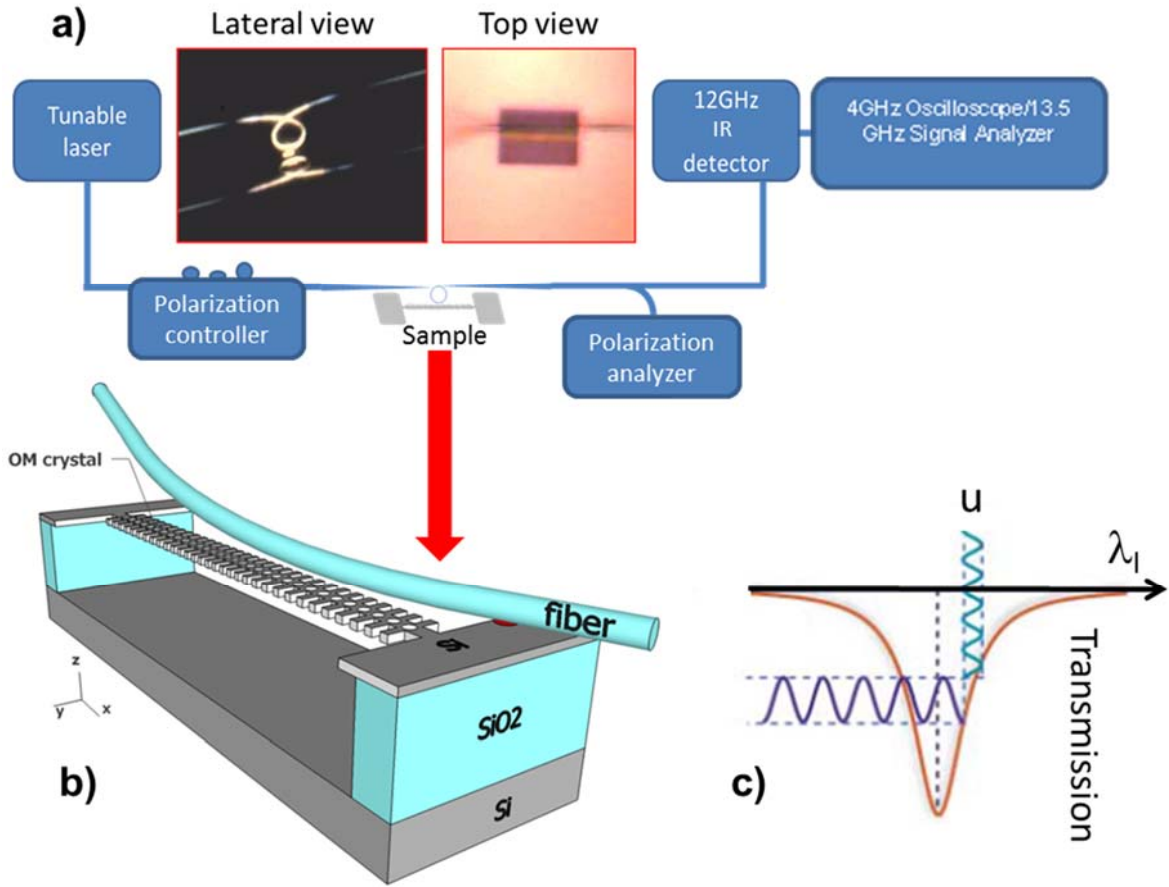

***Figure S6. a)***. *Sketch of the experimental setup to measure the optical and mechanical properties of the OM devices. The sample size has been greatly increased for clarity. The top left photo shows a lateral view of the real microlooped tapered fiber close to the sample, where the fiber can be seen reflected on the sample. The top right photo shows a top view of the tapered fiber placed parallel with the OM structure and in contact with one of the edges of the etched frame.* ***b)***. *Relative positioning of the tapered fiber and the OM photonic crystal. The leaning point of the fiber is highlighted in red. The fiber is placed close enough to the central part of the OM photonic crystal to excite efficiently its localized photonic modes.* ***c)*** *Scheme of the transduction principle.*

### S3.1. Tapered fiber characteristics and fabrication procedure

The experiments are carried out with tapered optical fibers having diameters in the smallest section of about 1.8μm (Fig. S7a), which is commensurate with the wavelength of interest (around 1.5 μm) and ensures an evanescent field tail of several hundreds of nanometers.

For the fiber fabrication, we used a home-made setup in which a SMF-28 optical fiber is stretched in a controlled way using two motorized stages. The central part of the fiber is placed in a microheater where the temperature is about 1180°C (2).

The fiber transmission at a wavelength of 1.5 μm is monitored during the pulling procedure (Fig. S7b). The signal is subjected to a short time Fast Fourier Transform(FFT) algorithm, so that the frequency components associated to inference between different supported modes are measured (Fig. S7c). The single mode configuration is achieved when all those frequency components disappear (Fig. S7d).

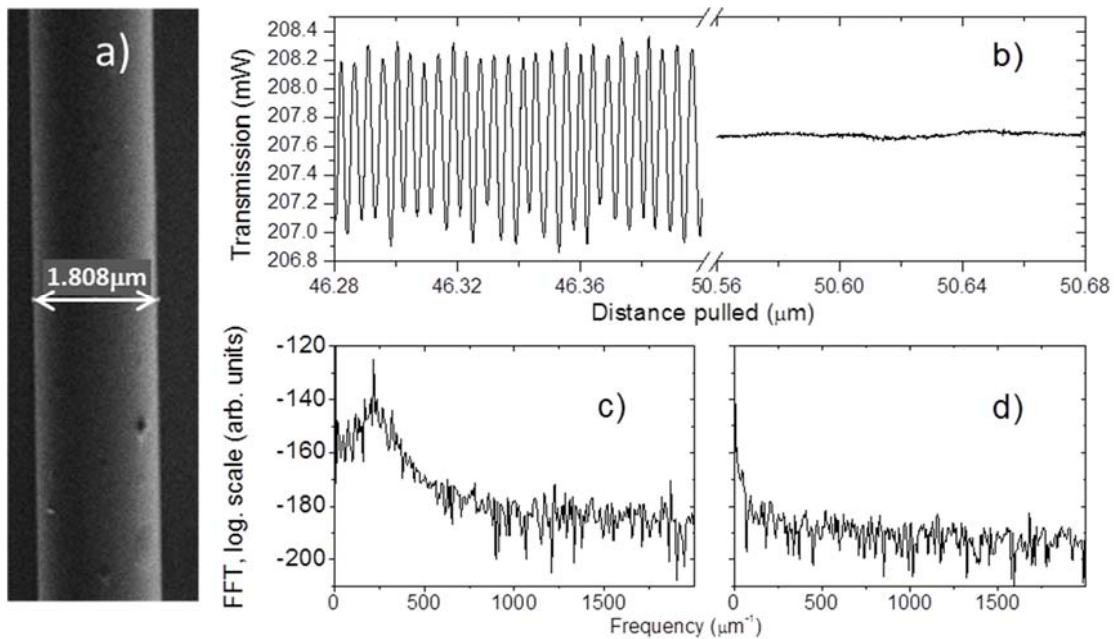

***Figure S7. a)*** *SEM image of the thinnest part of a tapered fiber.* ***b)*** *Transition from multimode to monomode while pulling the two extremes of the fiber as seen in temporal scale.* ***c)*** *FFT of the*

*transmitted signal before the transition to single mode. **d)** FFT of the transmitted signal after the transition to single mode.*

Using two rotating fiber clamps, the tapered fiber is twisted twice around itself. The two fiber ends are gently brought closer over several hundreds of micrometers so that a looped structure forms in the tapered region. The two fiber ends are afterwards pulled apart to reduce the loop size down to a few tens of μm. In that process, the two parts of the fiber at the loop closing point (upper part of the loop on the top left photo Fig. S7) slide smoothly in opposite senses. The micro-looped shape provides functionalities similar to those of dimpled fibers.

For the fiber loop photonic structure, the dispersion relation is linear and the group refractive index ($n_g$) is equivalent to the effective refractive index ($n_{eff}$). We have calculated $n_{eff}$ using the Beam Propagation Method taking the material refractive index of the cladding of the initial fiber to be n=1.468 at $\lambda_l$=1515nm, resulting in a value of $n_{eff}$=1.373.

### S4. Thermo-optic/Free-Carrier-Dispersion Self-Pulsing

In this section, we address self-pulsing (SP) limit-cycles in optical resonators. Those are stable periodic regimes established in the cavity that are associated to the dynamic competition among several optical non-linear mechanisms. When a SP limit-cycle is active, light within the cavity gets modulates in a strongly anharmonic way, creating an "optical frequency comb" with multiple peaks spectrally located at integers of main SP frequency ($\nu_{SP}$). The dynamic behavior of the intracavity photon number is then equivalent to what could be reached by externally modulating the laser input, although in this case the optical modes are pumped with a continuous wave source.

We report here on a SP mechanism resulting from the interplay between thermo-optic (TO) effects and free carrier dispersion (FCD) (6) within the OM crystal. In the case of FCD, the excess of free-carriers leads to a reduction of the material refractive index and therefore a blue-shift of the optical cavity mode. On the other hand, the TO effect results in an increase of the refractive index of the material with increasing temperature. Since the main source of heating is Free-Carrier Absorption (FCA), the dynamics of free-carrier density (*N*) and the temperature increase (*ΔT)* are linked and can be described by a system of coupled rate equations (6):

$$\dot{N} = -\frac{1}{\tau_{FC}}N + \beta\left(\frac{hc^3}{n^2\lambda_o V_o^2}\right)n_o^2 \quad \text{(S4a)},$$

$$\dot{\Delta T} = -\frac{1}{\tau_T}\Delta T + \alpha_{FC}Nn_o \quad \text{(S4b)}$$

where $V_o$ is the optical mode volume, $\lambda_o$ is the cavity resonance wavelength at room temperature and $\alpha_{FC}$ is defined as the rate of temperature increase per photon and unit free-carrier density. In the FCD equation (equation S4a) we consider a Two-Photon Absorption (TPA) generation term, where $\beta$ is the tabulated TPA coefficient and a surface recombination term governed by a characteristic lifetime $\tau_{FC}$. The TO effect (equation S4b) reflects the balance between the fraction of photons that are absorbed and transformed into heat due to FCA and the heat dissipated to the surroundings of the cavity volume, which is governed by a characteristic lifetime $\tau_T$. The generation terms of Equations S4 depend on $n$, which can be written as $n = n_o \dfrac{\Delta\lambda_o^2}{4(\lambda_l - \lambda_r)^2 + \Delta\lambda_o^2}$, where $n = n_o = 2 P_l \kappa_e \lambda_o / \kappa^2 hc$ in perfect resonance. $P_l$ and $\lambda_l$ are the laser power and wavelength, respectively; $\kappa_e$ and $\kappa$ and are the extrinsic and overall optical damping rates, respectively, the latter determining the cavity resonant linewidth ($\Delta\lambda_o = \lambda_o^2 \kappa / 2\pi$). Importantly, $\lambda_r \approx \lambda_o - \dfrac{\partial \lambda_r}{\partial N} N + \dfrac{\partial \lambda_r}{\partial T} \Delta T$ is the cavity resonant wavelength including first order nonlinear effects. Since $\kappa^{-1}$ is much smaller than $\tau_{FC}$ and $\tau_t$ it is possible to consider that the optical cavity responds adiabatically to the FCD and TO induced refractive index changes.

Experimentally, we have access to both laser parameters ($P_l$ and $\lambda_l$), are included in the dynamic equations implicitly in $n$, thus impacting the generation rates.

Figure S8 plots the simulated phase portraits {N, ΔT} for specific values of $\lambda_l$. The trajectories (red curves) are calculated by imposing initial conditions in such a way that the cavity is red-detuned with respect to the laser by an amount equal to $\Delta\lambda_o$. For clarity, we have also included the nullclines of the system, defined as the curves where either $\dot{N} = 0$ (vertical flow, green curves) or $\Delta\dot{T} = 0$ (horizontal flow, blue curves). The non-linear system of differential equations described by Equations S4a and S4b has generally a stable fixed point (solid circle in Fig. S8a), where both time derivatives are zero and the eigenvalues of the Jacobian matrix of the linearized system have a negative real part. Thus, if driven away from that point, the system settles down to equilibrium through exponentially damped oscillations. The system has a second fixed point (open circle), which in this case is unstable. When $\lambda_l$ reaches a specific value $\lambda_{l,th}$, the stable fixed point undergoes a supercritical Hopf bifurcation (red curve of Fig. S8b). Just at the bifurcation the fixed point is still stable, although it is a weak one, and, if found in its vicinity, the system decays to it only algebraically. Above $\lambda_{l,th}$ the fixed point becomes unstable and it is surrounded by a stable SP limit-cycle (Fig. S8c), where the system oscillates periodically at a frequency denoted by $\nu_{SP}$. The total time required to complete the {N, ΔT} limit-cycle is thus $1/\nu_{SP}$, although is not necessarily drawn at a constant pace. The SP regime is robust over a wide spectral range, just increasing the average cavity temperature around which the system is oscillating. Since the decay rate of the temperature difference ΔT in Equation S4b depends on ΔT itself, $\nu_{SP}$ can increase continuously by increasing the average amount of heat in the cavity. This is simply achieved by increasing $\lambda_l$, which increases the time-averaged $n$.

The SP limit-cycle is robust along a wide spectral range, and is only wiped out when the second fixed point gets close enough to deflect down the trajectory of the system (Fig. S8d), the cavity then relaxes down to the initial "cold" situation at ΔT=0, where the cavity resonance is far away from the laser line.

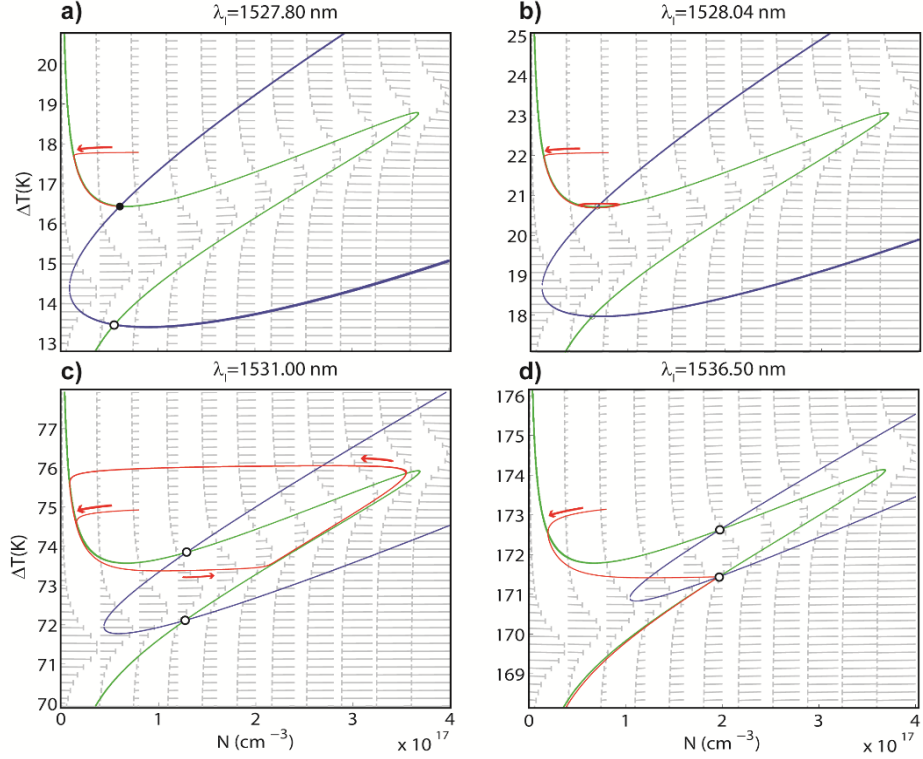

***Figure S8.*** *Phase portraits as a function of the free carrier density and temperature increment for $\lambda_l$ below the Hopf bifurcation (panel a)), at the bifurcation (panel b) and above the bifurcation (panels c) and d)). The nullclines of Equations S4a and S4b are plotted in green ($\Delta\dot{T}=0$) and blue ($\dot{N}=0$) respectively, while the system trajectories are in red. Grey arrows depict the temporal derivatives of $\Delta T$ and $N$ in each point of the phase space.*

### S5. Phonon lasing through Self-Pulsing pumping

As a result of the SP limit-cycle, the radiation pressure optical force ($F_o$) is modulated in the same way as the intracavity photon number, since they are related in a linear way, i.e., $F_o = \hbar g_{o,OM} n$. Neglecting the Langevin force, the mechanical modes of the nanobeam can be described as damped linear harmonic oscillators driven by the anharmonic force:

$$m_{eff}\ddot{u}_1 + m_{eff}\frac{\Omega_m}{Q_{m,i}}\dot{u}_1 + k_{eff}u_1 = F_o \qquad (S5)$$

, where $u_1$ is the generalized coordinate for the displacement of the mechanical mode and $m_{eff}$, $k_{eff}$ and $\Omega_m$ are its effective mass, spring constant and eigenfrequency, respectively. Finally, the nonlinear resonance position has to include now the effect of the mechanical motion when evaluating $n_o$, i.e., $\lambda_r \approx \lambda_o - \frac{\partial \lambda_r}{\partial N}N + \frac{\partial \lambda_r}{\partial T}\Delta T + \frac{\lambda_o^2 g_{o,OM}}{2\pi c}u_1$. Importantly, the response of $n$ to the mechanical deformation is also adiabatic since $\kappa$ is few orders of magnitude smaller than $\Omega_m$.

Equations S4 and S5 describe the dynamics of a four dimensional ($N$, $\Delta T$, $u_1$, $\dot{u}_1$) non-linear system that is coupled through $n$. Much more complex trajectories than those derived from isolated Equations S4 can now arise, opening the way to chaotic trajectories within a basin (8) of attraction in the phase space. One of the consequences of the coupling of the two systems is that self-sustained mechanical motion is achieved if one of the low harmonics of the SP main peak at $v_{SP}$ is resonant with the mechanical oscillations ($M v_{SP} = \Omega_m$ $M v_{sp} = \Omega_m$, where $M \in \mathbb{Z}$). In fact, the coherence of the mechanical oscillation is maintained since the mechanical mode lifetime is much longer than $1/M v_{SP}$

Fig. S9 shows the experimental RF spectra for the case of a mechanical mode at $\Omega_m$=54 MHz $M$=1 (black and gray curves) and $M$=2 (green curve). In the latter case, the mechanical oscillation at frequency $\Omega_m$ is superimposed on a SP trace at frequency $\Omega_m/2$. Although the black and gray curves are obtained at two different values of $\lambda_l$, the signal is locked at the same frequency $\Omega_m$. Here it becomes evident that the mechanical oscillator is not only pumped resonantly, but also that the large amplitude of the coherent mechanical motion acts as a feedback that stabilizes and entrains the SP and the mechanical oscillator. Since $\Omega_m$ is much more robust than $v_{sp}$, the SP mechanism adapts its frequency to the mechanical one. When the resonant condition with the mechanical oscillation is not fulfilled (red curve of Fig. S9a) the RF peaks in a frequency-unlocked region are inhomogenously broadened in frequency because the integration time is greater than the unlocked frequency oscillation.

The experimental features of this coupled system, can be well reproduced with the couple set of equations S4 and S5, the only free parameters being $\tau_{FCA}$ and $\tau_t$. Those are extracted by fitting the temporal trace of the transmitted signal in an unlocked configuration using Equations S4 (7). The fitting parameters used to reproduce the frequency-unlocked cases are $\tau_T$=0.5 [µs], $\tau_{FC}$=0.5 [ns] and $\alpha_{FC}$=4x10$^{-13}$ [K cm$^3$ s$^{-1}$], while the initial conditions verify that $\frac{\partial \lambda_r}{\partial \Delta T}\Delta T(0) = \lambda_l - \lambda_o$. TO and FCD coefficients were independently calculated by assuming that the observed wavelength shift is only associated to an average change in the Si refractive index within the region overlapping with the electromagnetic fields and using tabulated values for its dependence with temperature and free-carrier density. This procedure lead to the following values: $\frac{\partial \lambda_r}{\partial N}$=7x10$^{-19}$ [nm cm$^3$] and $\frac{\partial \lambda_r}{\partial \Delta T}$=6x10$^{-2}$ [nm K$^{-1}$].

Fig. S9b illustrates the simulated limit-cycle trajectories followed by the system under equivalent conditions as those of the curves of Fig. S9a. Coherent mechanical oscillations of nm amplitude are obtained in the $M$=1 and $M$=2 cases, while the red curve is flat in the u-axis.

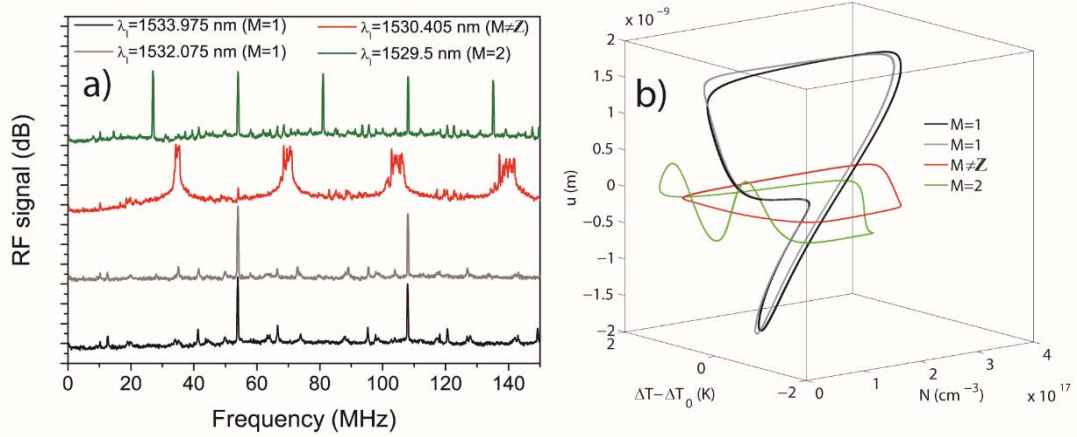

*Figure S9.* a) Experimental RF spectrum at different values of $\lambda_l$. The black and gray curves correspond to M=1 situations, while the green spectrum corresponds to M=2. The red curve corresponds to an unlocked situation. b) Simulated phase portraits calculated using our coupled model for equivalent situations to those of panel a).

## S6. Modelling of hysteresis and chaos (only one mechanical oscillator at $\Omega_m$=54MHz).

We have numerically solved the system of four non-linear equations composed by Eqs. S4 and S5. The simulations are done sequentially for increasing (decreasing) values of $\lambda_l$ using initial conditions provided by one point ($N$, $\Delta T$, $u_1$, $\dot{u}_1$) of the stable state calculated for the preceding $\lambda_l$ value. The obtained results are summarized in Fig. S10, where we plot the FFT of the transmitted signal. The model predicts, on the one hand, bi-stability and hysteresis between bi-dimensional SP and tetra-dimensional SP/phonon lasing, thus displaying some of the features reported in the main text.

The model also accounts for the onsetting of chaos on ($N$, $\Delta T$), while the mechanical mode oscillates coherently (sharp peaks appear at $\Omega_m$=54MHz and its harmonics). However the predicted route is much smoother than the experimental and through subsequent period doubling bifurcations instead of through an abrupt transition. It is worth noting that, again here, the period doubling dynamics is only present in ($N$, $\Delta T$), while the mechanical mode oscillates in a coherent fashion.

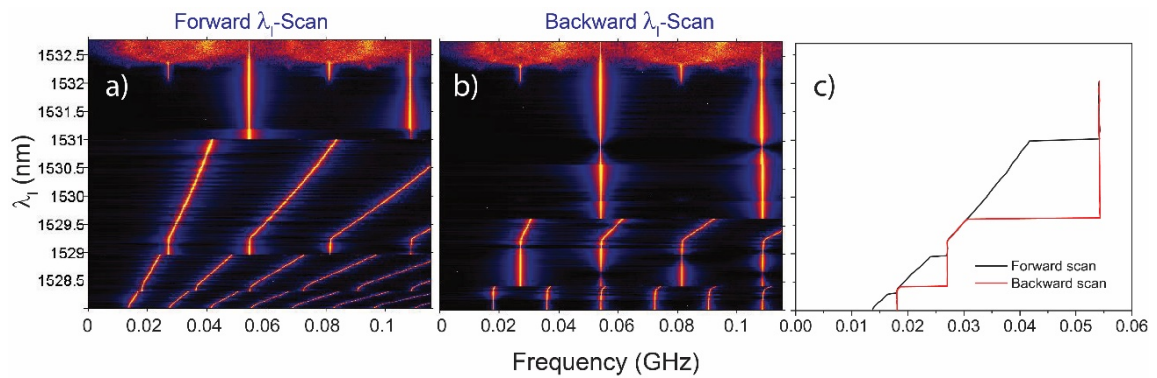

*Figure S10. Panels **a)-b)**. Color contour plot of the simulated RF spectrum of the transmitted signal as a function of $\lambda_l$ obtained for $n_o=10^5$. Panel **a)** correspond to a forward $\lambda_l$-scan while panel **b)** to a backward $\lambda_l$-scans. Panel **c)** plots the spectral position of the first RF harmonic.*

On Fig. S11 we show the simulated trajectories associated to a tetra-dimensional SP/phonon lasing (left panel) and to chaos.

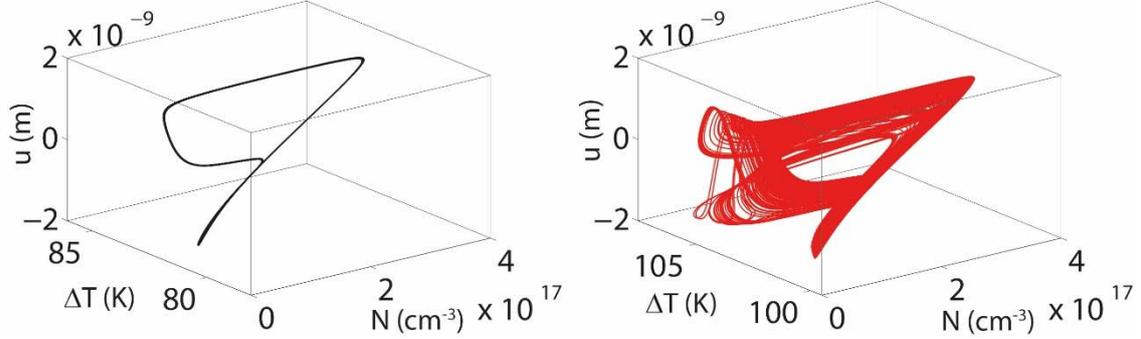

*Figure S11. Numerical Simulations of the phase-space trajectories associated to a tetra-dimensional SP/phonon lasing (left panel) and within a chaotic regime (right panel).*

The model described by Eqs. S4 and S5 captures most of the features described along the main text. However, it does not incorporate three main experimental observations, namely: i) the period doubling bifurcation route between M=2 and M=1, ii) the presence of bistability and hysteresis between tetra-dimensional SP/phonon lasing and chaos and iii) the specific peak structure of the chaotic RF signal. The first observation remains a subject for further studies, since we predict much smaller high order plateaus that what measured experimentally, which eventually prevents reaching a similar situation to that reported in Fig. 4c. In the following section we address the former observations.

### S7. Modelling of the chaotic dynamics with two mechanical oscillators (at $\Omega_m$=54MHz and at $\Omega_m$'=5MHz).

As explained briefly in the main text, most of the sharp peaks that are present in the chaotic RF spectrum reported in Fig. 4 of the main text are associated either to: i) the $\Omega_m$ mode (main peak + harmonics) that was lasing in the plateau, similarly to what reported in the simulation of the previous section. The remaining peaks are associated to the activation of the fundamental out-of-plane flexural mode described before ($\Omega_m$'=5 MHz), providing peaks at $\Omega_m$', at its harmonics and at sidebands of the $\Omega_m$ associated peaks.

In order to account for that mode in the model, we have included a second harmonic oscillator in addition to Eqs. S4 and S5. The force driving the second oscillator is very likely radiation pressure ($F_o$'), because that particular mode displays the stronger transduced signal below the threshold for SP, where thermal Langevin forces are dominant. We rule out pure mechanical non-linear interactions because, if they were to be accounted for, other mechanical modes should also be active when the system is in a plateau, which is not the case. The additional equation thus reads:

$$m_{eff,2}\ddot{u}_2 + m_{eff,2}\frac{\Omega_m'}{Q_{m,i,2}}\dot{u}_2 + k_{eff,2}u_2 = F_o' \qquad (S6)$$

The $g_{o,OM}$ value for this specific mode has been taken by scaling the $g_{o,OM}$ value corresponding to the $\Omega_m$ =54MHz mode by the RF signal ratio between the two modes. The obtained results are very similar to those of the previous section. Within the M=1 plateau, the second harmonic oscillator is off (see Fig. S12b). However, when the system enters into the chaotic regime, it gets activated as a consequence of the chaotic dynamics of $F_o'$. Through OM coupling its oscillations modulate $\lambda_r$. As a result the FFT of the transmitted signal displays sharp peaks at $\Omega_m$ and $\Omega_m'$ (see top panel of Fig. S12a). It is also possible to distinguish sidebands at $\Omega_m \pm \Omega_m'$. The first harmonic oscillator oscillates coherently also in the chaotic regime, which again is only present in (N, $\Delta T$). In Fig. S13, we report temporal series spanning 1μs of the different magnitudes described by the non-linear system of equations S4, S5 and S6 (we have skipped deformation velocities). In the bottom panel of Fig. S13 we plot the resulting transmitted signal.

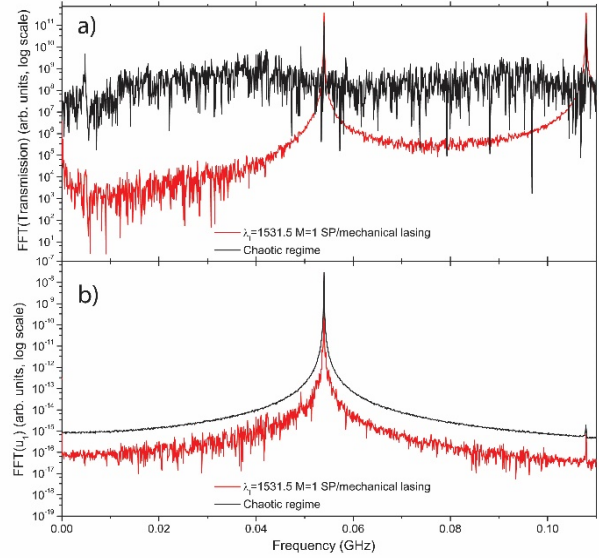

**Figure S12.** FFT of the simulated transmitted signal (black) and deformation of the first oscillator (red) below and within the chaotic regime (panels a) and b), respectively).

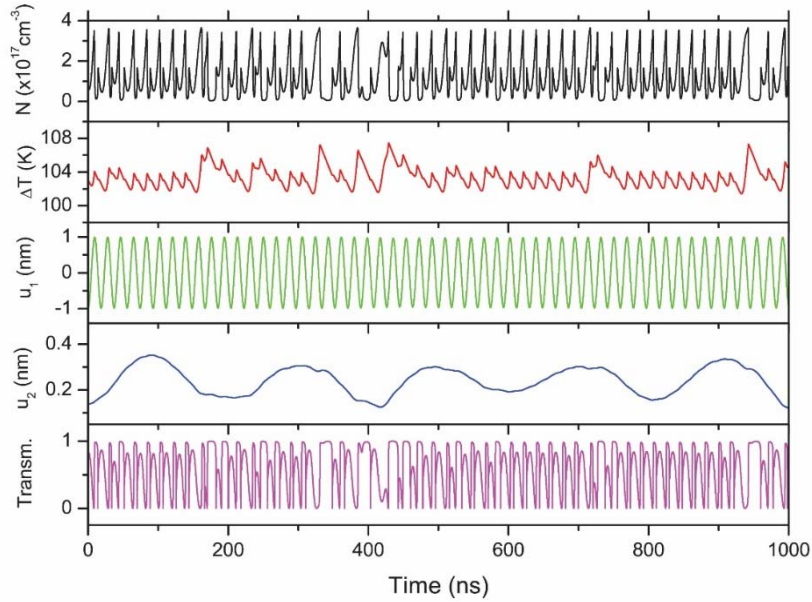

***Figure S13.*** *Temporal series of the different magnitudes extracted by the model. From top to bottom: N, ΔT, $u_1$, $u_2$ and Transmission).*

## S8. Application of Rosenstein algorithm to the experimental temporal series.

We have used the Rosenstein method for calculating the largest Lyapunov exponent (LLE) from our experimental series (9). The output is the temporal evolution of the average divergence of initially close state-space trajectories. The method follows directly from the definition of LLE and is accurate because it uses the whole available set of data. Although it is suitable for relatively small data sets, the temporal series has to be well resolved and span over a temporal range that should be several times greater than the predictability horizon of the system. Otherwise, the results will not be trustworthy and the extracted LLE values will be orders of magnitude greater than the real ones, i.e. the trajectories coming from as-close –as-possible initial states will almost immediately diverge up to the scale of the attractor volume. Indeed, this was our case when applying the Rosenstein algorithm to short chaotic time series (few hundreds of ns), since, given a reference state on the time series, it was not possible to find a good as-close-as-possible state in the remaining set of data.

Another critical input value is the embedding dimension of the system (m). It is imperative to evaluate the algorithm using different values for m. In Fig. S14 it is apparent that satisfactory results are obtained only when m is, at least, six, which is indeed the dimension of the system. This is due to the fact that chaotic systems are effectively stochastic when embedded in a phase space that is too small to accommodate the true dynamics.

Finally, it is worth noting that the superposition of two sinusoids (for instance, two harmonic oscillators) with incommensurate frequencies is a non-chaotic system, but is quasiperiodic and deterministic. Although the time series would look chaotic, applying the Rosenstein algorithm to such a signal would lead to a flat divergence curve (9), leading to a null LLE. In our case the divergence is clearly growing with time, which is a clear signature of chaos.

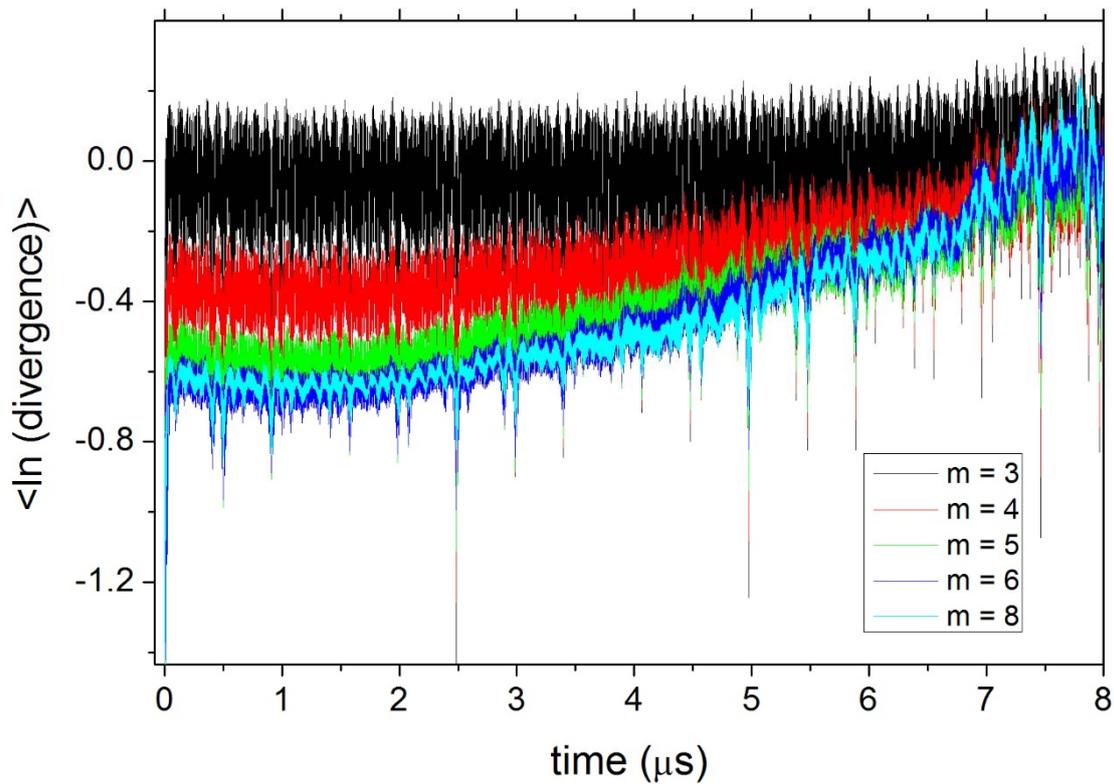

***Figure S14.*** *Time evolution of the <ln (divergence)> extracted from the experimental signals by applying Rosenstein algorithm for different values of the embedding dimension. The other input parameters are: delay=10ns; mean period=20ns. The total time register is 10µs long, acquired with a resolution of 10ps.*